\documentclass[a4paper,11pt,onecolumn]{article}
\usepackage{amsmath}
\usepackage{amsfonts}

\usepackage{amssymb}
\usepackage{graphicx}
\usepackage{bm}
\usepackage{geometry}
\usepackage{cite}

\usepackage[plainpages=false,hypertexnames=true,hyperindex=true,pagebackref=false,raiselinks=true]{hyperref}
\hypersetup{bookmarks,bookmarksnumbered,colorlinks,linkcolor=blue,breaklinks=true,linktoc=all}

\begin{document}

\title{The outcomes of measurements in the de Broglie-Bohm theory-II}
\author{F. Lalo\"{e}\thanks{%
laloe@lkb.ens.fr}\; and G. Tastevin \\
Laboratoire Kastler Brossel, ENS-Universit\'e PSL,\\
CNRS, Sorbonne Universit\'e, Coll\`ege de France,\\
24 rue Lhomond 75005\ Paris, France}
\date{\today}
\maketitle

\begin{abstract}
Part I of this article discussed the quantum measurement process within the de Broglie-Bohm theory. In the experiment considered, the outcome of the measurements was primarily determined by the initial
Bohmian positions within the measurement apparatus.
This is nevertheless not always the case. Part II is a short addendum with a more general discussion: cases where the measurement results reveal the values of
the positions attached to the measured system, the measurement apparatus, or both, are considered.
\end{abstract}

  In Ref. \cite{GT-FL-1}, we discussed the quantum measurement process within the de Broglie-Bohm (dBB) theory in order to emphasize that, in some cases,
the outcome of the measurements was primarily determined by the initial
Bohmian positions within the measurement apparatus M; contrary to what intuition suggests, the positions attached to the measured system S then play no role in
the determination of the outcome. This is nevertheless not always the case: in other conditions, the opposite is true, and what determines the results of measurement are the positions of S, not those of M; as in classical physics, performing a measurement then reveals a pre-existing property (within the dBB theory) of S. See also for instance the discussion of Ref. \cite{Aleshin-et-al}. The present text is an addendum to Ref. \cite{GT-FL-1}  containing a more general discussion of the measurement problem,  and distinguishing between cases where the measurement results are determined by
the Bohmian positions of S, or M, or both. This can be seen as a discussion
of the mechanism of contextuality within the dBB theory.

\begin{figure}[h]
\centering
\includegraphics[trim = 60mm 55mm 65mm 40mm, clip,width=9cm]{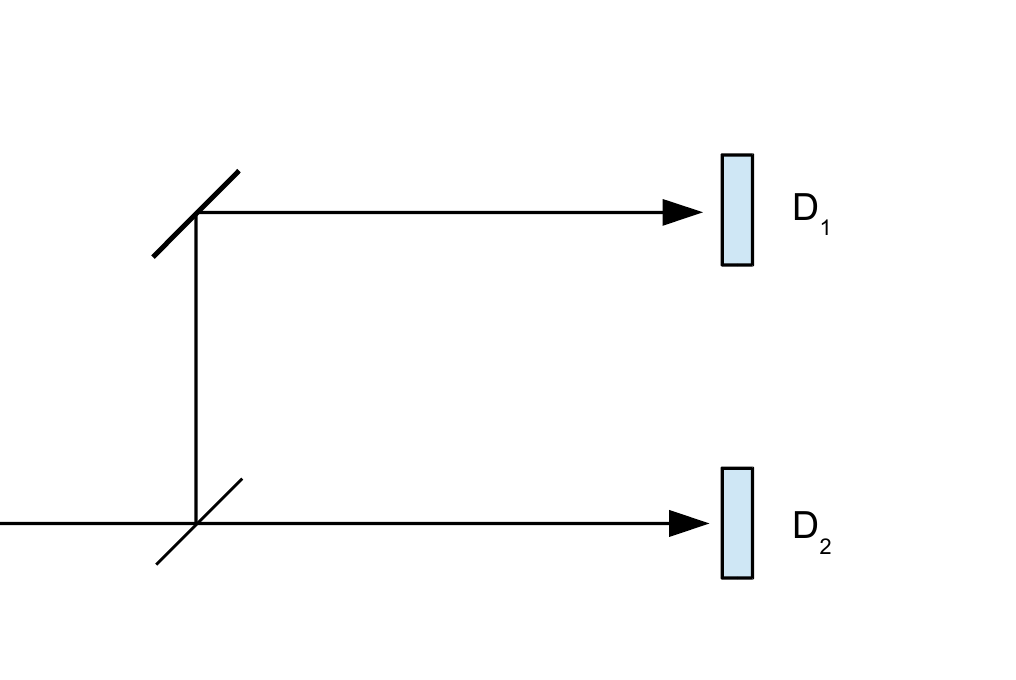}
\caption{A very simple experiment where the Schr\"{o}dinger wave of a single particle is split into two components, which are then sent on two detectors D$_1$ and D$_2$.}
\label{f1}
\end{figure}

\textbf 1 - A first, very simple, example is illustrated in figure~\ref{f1}: a beam splitter separates the quantum wave of a single particle into two components, which are then directed towards two detectors D$_1$ and D$_2$. The particles may be for instance metastable Helium atoms detected with microchannel plates \cite{Wesbrook}, while the atomic beam splitter can be obtained optically with near-resonant lasers \cite{Adams}. When a metastable atom reaches the entrance of a microchannel, an electron is ejected from the solid surface which, under the effect of an applied electric field, triggers an electronic cascade inside the channel. This amplification effect eventually leads to  in a macroscopic electric current, and to a  click of the detector. If the atoms are sent one by one into the system,  in each realization of the experiment only one of detectors clicks, while the other remains silent.

In this simple experiment, which detector clicks is determined by the trajectory of the particle in the optical device. This is because the Bohmian trajectory is continuous and cannot jump from one path to the other; when it reaches one of the detectors, the single particle trajectory determines which component of the many particle state vector is active and contains the cascade process, and which remains an \textquotedblleft empty wave\textquotedblright . Mathematically, the state vector describing the whole system of the atom and the two detectors contains no component where the atom is in one arm while the variables of the detector in the other arm are subject to the cascade effect; the Bohmian positions cannot explore regions of the configuration space where the wave function vanishes. As a consequence, in this case, and as one could naively expect, the outcome of the measurement is indeed determined by the trajectory of the measured atom, that is by its position within its wave packet just after the crossing of the beam splitter.

A similar situation occurs when a beam of particles
is sent onto a cloud chamber in order to detect the trajectories of particles. When the wave packet of a particle enters the cloud chamber and initiates a track by a cascade of condensation effects, it is the Bohmian position of the particle that determines the location of the start of the track, and therefore its continuation in the chamber, as discussed for instance by Darwin and Mott \cite{Mott}. Again, in this case the result of measurement gives an information on the position of the particle before measurement; the Bohmian positions of the particles in the cloud play no essential role.

\pagebreak

\textbf 2 - Maybe the example of quantum measurement most frequently discussed in textbooks remains the Stern-Gerlach experiment -- even if in practice this category  experiment is rarely performed in laboratories. This case is special since no cascade or amplification effect is involved, so that no single particle detection really occurs; what is done is to repeat the experiment a very large number of times, until visible spots of atoms are created on the detection plate. This allows the observation of   the quantization of the spin of the Ag atoms, without any individual observation.

Reference \cite{Bricmont-2016} provides an interesting discussion of the Stern-Gerlach experiment within dBB theory. It emphasizes the fact that the trajectory of the each atom deviates upwards if its  initial position is initially in the upper part of the wave packet,  downwards if it is initially in the lower part\footnote{Relativistic considerations show that the Bohmian velocity includes a rotational spin-dependent term (Gordon term) \cite{Holland-Philippidis, Das, Wilkes} that is not taken into account in this discussion. The Gordon term slightly modifies the trajectories \cite{Holland-Philippidis}, but does not change the main conclusions concerning the strong contextuality of the measurement results.}. The surprising result is that this situation does not change when the direction of the gradient of the magnetic field is inverted, so that the relation between values of the spin and trajectory deviations is swapped. As a consequence, in an experimental situation a given initial position of the atom determines a value $+ \hbar / 2$ of the spin, in another situation an opposite value $- \hbar / 2$. In other words, the initial atomic position determines the result of measurement (the value of one component of its spin), but in a way that depends on the design of the apparatus; the result is contextual.

\textbf 3 - Reference \cite{GT-FL-1} discusses another experiment, an \textquotedblleft optical version of the Stern-Gerlach experiment\textquotedblright , where two light beams with opposite circular polarizations create spin dependent recoil effects on the atoms. In this case, the two spin states of the atom occupy initially the same region of space; they start to separate only under the effect of the interaction with the measurement system. Two competing effects then take place: on one hand, the overlap between the two wave functions of the atom associated with the two spin directions decreases as a function of time, but on the other hand the overlap between the wave functions of the measurement apparatus associated with the two results also starts to decrease. Because the measurement apparatus contains a large number $N$ of particles which, each, contribute to the decrease of the overlap, the $N$-particle wave functions become much more rapidly orthogonal in the configuration space of the apparatus than those of the atom. At the end, the choice of the result of measurement depends only on the initial positions of the particles inside the measurement apparatus, while the Bohmian position of the atom itself is practically irrelevant. It is the measurement apparatus that determines the result, and then drives the evolution of the position of the atom accordingly. The conclusion is therefore the exact opposite of that of the preceding paragraph, which is not contradictory since the experiment is different.

\enlargethispage{4mm}
One may notice that the Gordon term was not included in the calculations of Ref.~\cite{GT-FL-1}, which simplifies the equations. The discussion relies mostly on the motion of wave packets (not affected by perturbations of the Bohmian trajectories) and on the trajectories of the Bohmian positions of the measurement apparatus (which are not subject to a Gordon term). The conclusions of   Ref.~\cite{GT-FL-1} should therefore remain almost unchanged when this term is included.

\textbf 4 - One can also have intermediate situations, for instance if the system S first interacts with an intermediate quantum system S' containing $N'$ particles, which eventually interacts with a macroscopic measurement apparatus containing a very large number of particles. Depending on the dynamics of the interaction between S and S' and on the value of $N'$, the information carried by S' to the measurement apparatus may depend primarily on the position of the Bohmian variables of S, or of those of S'; the dynamics of interaction between S' and the measurement apparatus also plays a role. Therefore, in such an indirect measurement using an ancilla quantum system S', depending on the conditions the result of measurement may be determined by the initial positions of either S, S', or M, or be a compromise between different Bohmian variables.

\vspace{2mm}

In conclusion, what matters is the  spatial extent  (before the measurement) of  the parts of the total wave function that, under the effect of the interaction with the measurement apparatus M, will become entangled with orthogonal wave functions of S that are situated in different regions (of its configuration space). In other words, what matters is the spatial extent of the wave functions associated with the eigenvectors  of the measured observable of S.   If these eigenfunctions are already spatially separated, the result of measurement is determined by the Bohmian position of S; if they overlap spatially, the Bohmian position of M may play an important role, or even determine the measurement result as in \cite{GT-FL-1}.

In experiments where the position of a particle is directly measured, the result of measurement (the place at which the particle materializes, for instance in a cloud chamber) is indeed determined by the Bohmian trajectory of the position of the particle. This is what intuition predicts; it remains true when the correlations between the positions of several particles are measured, for example as in Ref. \cite{Kazemi}. By contrast, in experiment where another observable is measured, such as an internal variable of the particle (spin), the situation may be the opposite. In this case, the result of measurement may be a consequence of the initial state of the measurement apparatus (as defined by the initial position of its Bohmian positions).

Acknowledgment: The authors thank Siddhant Das for interesting comments and discussions.

\end{document}